\documentclass[%
 prb,
 jmp,%
 amsmath,amssymb,
preprint,%
author-year,%
showkeys,
superscriptaddress,
]{revtex4}

\usepackage{graphicx}
\usepackage{dcolumn}
\usepackage{bm}

\begin{document}


\title[]{Quasi-dark Mode in a Metamaterial for Analogous Electromagnetically-induced Transparency}

\author{V. T. T. Thuy}
\affiliation{Quantum Photonic Science Research Center and Department of Physics, Hanyang University, Seoul 133-791, Korea}
\author{N. T. Tung}
\affiliation{Quantum Photonic Science Research Center and Department of Physics, Hanyang University, Seoul 133-791, Korea}
\author{J. W. Park}
\affiliation{Quantum Photonic Science Research Center and Department of Physics, Hanyang University, Seoul 133-791, Korea}
\author{Y. H. Lu}
\affiliation{Quantum Photonic Science Research Center and Department of Physics, Hanyang University, Seoul 133-791, Korea}
\author{Y. P. Lee}
\email[Corresponding author. Electronic address: ]{yplee@hanyang.ac.kr}
\affiliation{Quantum Photonic Science Research Center and Department of Physics, Hanyang University, Seoul 133-791, Korea}
\author{J. Y. Rhee}
\affiliation{Department of Physics, Sungkyunkwan University, Suwon 440-746, Korea}
\date{\today}

\begin{abstract}
We study a planar metamaterial supporting electromagnetically-induced transparency (EIT)-like effect by exploiting the coupling between bright and quasi-dark eigenmodes. The specific design of such a metamaterial consists of a cut-wire (CW) and a single-gap split-ring resonator (SRR). From the numerical and the analytical results we demonstrate that the response of SRR, which is weakly excited by external electric field, is mitigated to be a quasi-dark eigenmode in the presence of strongly radiative CW. This result suggests more relaxed conditions for the realization of devices utilizing the EIT-like effects in metamaterial, and thereby widens the possibilities for many different structural implementations.

\end{abstract}

\pacs{Valid PACS appear here}


\maketitle

Electromagnetically-induced transparency (EIT) is a coherent process in which an opaque atomic medium is rendered to be transparent in a narrow spectral region within a broad absorption band through the quantum interference of pump and probe laser beams tuned at different transitions. \cite{1,2} That narrow and highly dispersive transmission results in a drastic reduction of the group velocity of light, \cite{3,4} which offers possibilities for many important applications in optical communications and quantum optics by enhancing nonlinearity. \cite{5,6,7} However, the experiments of quantum EIT require a special setup with extreme conditions, as well as are limited to certain quantum systems and specified spectral region. Recent discoveries of EIT-like effect in many classical systems, including coupled, microresonators \cite{8} waveguide side-coupled to resonators \cite{9} and metamaterials, \cite{10,11,12,13,14,15} have therefore opened new gateways to overcome those issues.

Up to now, experimental and theoretical studies of EIT-like effect in metamaterials have been based on two main approaches; the ``trapped mode" resonances, \cite{10} and the coupling between ``bright" and ``dark" eigenmodes. \cite{11,12,13,14} Interestingly, the analogy of EIT-like effect has been also demonstrated by the coupling of two resonators which are both resonantly excited by the external field, i.e., there is no intrinsically dark eigenmode. \cite{15} This structure is called modified split-ring resonator (SRR) structure, consisting of an asymmetric metallic SRR enclosed by a closed metallic ring. Their fundamental modes have identical frequencies but are strongly deviated in spectral qualities, which are assigned as dark and bright eigenmodes, respectively. The analogy of EIT has also been drawn and discussed in detail. Nonetheless, the real electromagnetic mechanism of this non-intrinsically dark mode has not been clarified. \cite{15}

In this letter, by applying the $RLC$-circuit model to the modified SRR (ring-SRR) structure, \cite{12,14,16} we propose a more straightforward metamaterial design in which the ring is replaced by a cut wire (CW). This replacement makes the structure more convenient for nanoscale fabrication while keeping the spectral behavior unchanged. By combining both numerical and analytical results, we demonstrate that, even though the SRR alone can act as a radiative component, it is mitigated to be ``quasi-dark" one in the presence of strongly radiative CW. On the other hand, our design exhibits a wide range of trade-off between group index and transmittance, which is highly important for applications.

The schematic of our structure (CW-SRR) and the equivalent $RLC$ circuit are presented in Figs. 1(a) and 1(b), respectively. The CW and the SRR are assumed to be gold, having plasma and collision frequencies of $\omega_{p}=1.367 \times 10^{17}$ s$^{-1}$ and $\omega_{c}=1.22\times10^{14}$ s$^{-1}$, respectively. \cite{13,17} They are periodically arranged in $x$ and $y$ directions with a periodicity of $1100$ nm. In our numerical simulation using CST Microwave Studio, the normal incidence light with \textbf{E}-field polarized along $y$ axis, and  \textbf{H}-field polarized along $x$ axis was employed. The group index $n_{g}$ is calculated from the retrieved effective refractive index $n$ \cite{18} via formula $n_{g}=\omega(dn/d\omega)+n$.

The simulated transmission spectra of structures of CW only and SRR only (SRR and CW are removed, respectively), and CW-SRR structure with a separation $d=100$ nm are presented in Figs. 2(a) $-$ 2(c), respectively. The fundamental mode of CW shows a dipole-like character with a high radiative loss, a broad bandwidth and a low $Q$-factor ($Q_{1}=1.6$); in contrast, the asymmetric SRR has $LC$-resonant properties with a lower radiative loss, a narrow bandwidth, and a higher $Q$-factor ($Q_{2} = 11.5$). \cite{11,13,19,20} The resonance frequency of CW is about 105 THz and that of the SRR is slightly higher. Interestingly, when these constituents are brought into the CW-SRR structure, the interference between two eigenmodes leads to a narrow and pronounced transmission window, together with a steep normal dispersion (II) lying between two closely spaced transmission dips (I, III). This effect is well-known as EIT-like effect. \cite{11,12,13,14,15} The positive group index $n_{g}$ exceeding 50, observed at the transmission window, [Fig. 2(d)] offers the possibility for a variety of slow-light applications.

In Fig. 1(b), the left and the right loops of circuit represent the CW and the SRR, respectively. Both of them are coherently driven by ac voltages to mimic the excitation of \textbf{E}-field on CW and SRR. This scheme is, therefore, distinguished from previous cases in which only the lower $Q$-factor one is driven, but the higher $Q$-factor one is being kept completely ``dark" from the direct excitation. \cite{12,14}

The charge amplitude $q_{0n}$ on capacitor $C_{n}$, its phase $\varphi_{n}$, and the real part $P_{n}$ of complex power $S_{n}$ in the $n$th loop ($n=$ 1 and 2) are first analytically calculated \cite{12,14,21} and then plotted in Figs. 3(a) and 3(b) with specified values of lumped elements, \cite{22} assuming that $v_{1}\approx v_{2}=Ve^{jwt}$. Due to the destructive interference of the oscillating currents in two loops, \cite{21} $q_{01}$ show a narrow dip at II$^\prime$ and two peaks at I$^\prime$ and III$^\prime$. The charge amplitude $q_{02}$, in contrast, exhibits a very strong peak in region II$^\prime$. These results, obtained from the $RLC$-circuit model, are consistent with both the monitored field $E_{y}$ presented in Figs. 3(c) $-$ 3(e), and the probed electric strength at the end of CW and the gap of SRR (not shown).

It seems to be counter-intuitive that the maximum of local field strength $E_{y,max}$ at the transmission window is about two times higher than that at stop bands. Nonetheless, the transmission window can be well elucidated if we consider the following facts. (1) The current in the strongly radiative component, CW, which contributes primarily to the electromagnetic response of the metamaterial, is suppressed ($q_{01}\approx0$). The radiation loss by this ($P_{1}$) is thus minimized. (2) Charge $q_{2}$ oscillates with a significant amplitude but \emph{in phase} with the driving voltage, i.e., $\varphi_{2}\approx0$, as seen in Fig. 3(a). Thus, the complex power $S_{2}=i_{2}^{*}v_{2}\sim i_{2}^{*}v=-jwq_{02}Ve^{-j\varphi_{2}}$ yields a small real power $P_{2}=$ Re$(S_{2})$, and a large reactive power Im$(S_{2})$. This case is opposite to the case of non-coupled, single $RLC$ circuit where charge and driving voltage are out of phase by $90^{0}$ at resonance: the real power is dominant, and the reactive power is zero. Overall, the total real losses $P=P_{1}+P_{2}$, presented in Fig. 3(b), exhibits a dip in region II$^\prime$. Equivalently, we can explain why the transmission can take place even though there exist strong current, charge and local \textbf{E} field by SRR. It should be also noticed here that, while SRR is weakly excited by external \textbf{E} field, it is neither completely dark nor strongly radiative (in comparison with CW). Therefore we regard this higher $Q$-factor component as a ``quasi-dark" one in the presence of strongly radiative CW. This concept of quasi-dark mode indeed loosens the practical conditions for EIT-like effects, suggesting the possibility for many additional matamaterial designs.

The trade-off between transmittance and bandwidth, and group index of the metamaterial is illustrated in Fig. 4. By modifying the separation $d$ from 100 to 480 nm, the maximum transmittance is reduced from 0.8 to 0.3 and the group index is increased from 50 to 220. The transmission and the group index of the ring-SRR structure is also presented in Figs. 2(c) and 2(d), respectively, for comparison. While the maximum transmittance of 0.87 is obtained by this structure, it comes at the price of a much lower group index of 30, and the tunability of those quantities are limited when the inner SRR is displaced within the boundary of outer ring of the structure, as used in Ref. 15. The aforementioned trade-off is often characterized by the well-known delay-bandwidth product which is important in communications and quantum optics. \cite{5,7} We introduce here another figure of merit (FOM), namely, transmittance-delay product (TDP). The TDP is crucial for nonlinear applications where the enhancement of energy density of propagating light in the medium is the most concerned. This FOM peaks at $d\approx450$ nm where the high energy density of the slow light in the medium is best promoted. This trade-off, together with a small shift of transmission peak according to $d$, will be discussed in detail in a separated paper.

In conclusion, we have studied the EIT-like effect by the coupling between bight and quasi-dark eigenmodes in a metamaterial consisting of one CW and one SRR in the unit cell. From the simulated and the analytical results, the electromagnetic mechanism for the quasi-dark mode of SRR have been discussed and elucidated. On the other hand, the simplicity of our proposed structure makes it more convenient for fabrication and miniaturization in the high frequency regime. The tunability of transmittance and effective group index, and the trade-off between them are also studied and characterized by the FOM. This study not only provides a promising candidate for slow-light applications but also suggests the new possibilities for EIT-like effects which are less restricted to the perfectly dark component.

V.T.T.T. would like to thank Dr. W. H. Jang at Korea Communication Commission for his help in the numerical simulation, Dr. A. A. Bettiol and Mr. S. Y. Chiam for sharing their knowledge. This work was supported by the NRF through the Quantum Photonic Science Research Center at Hanyang University, Korea. This work was also supported by NRF grant funded by Korean Government (MEST) (Grant. No. KRF-2008-005-J00703).

\clearpage
\begin{center}
\textbf{Figure captions}
\end{center}

FIG. 1. (Color online) (a) Schematic of the CW-SRR metamaterial. CW has dimensions of $l_{1}=1025$ nm and $w_{1}=160$ nm. SRR has sizes of $l_{2}=l_{3}=380$ nm, $w_{2}=80$ nm and $g=160$ nm. The gold thickness is $t=80$ nm, and the separation between CW and SRR is $d$. (b) Equivalent $RLC$-circuit model.\\

FIG. 2. (Color online) Transmission spectra (blue solid curves) of (a) CW only and (b) SRR only. (c) Transmission spectrum and (d) retrieved group index (blue solid curves) of CW-SRR structure. The black dashed curves are the spectra of the corresponding structures in which CW is replaced by ring.\\

FIG. 3. (a) Charge amplitudes $q_{01}$, $q_{02}$, their corresponding phase shifts $\varphi_{1}$, $\varphi_{2}$. (b) Dissipated power on loop $P_{1}$ or $P_{2}$ and on entire circuit $P$. Simulated electric field E$_{y}$ at (c) left stop band (I), (d) transmission window (II) and (e) right stop band (III). Also see Fig. 2(c) for I, II, and III. ($+/-$) represent the signs of accumulated charges.\\

FIG. 4. Dependence of (a) the transmission, and (b) the group index and the FOM on the separation between CW and SRR.

\clearpage
\begin{figure}
\includegraphics[width=10cm,angle=0]{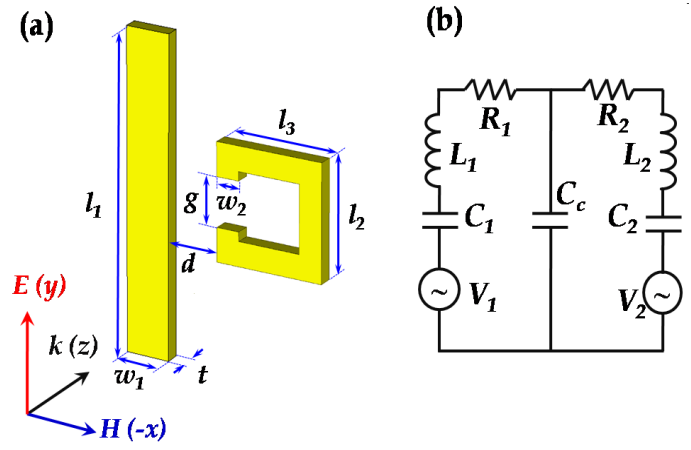}
\caption{Thuy \emph{et al.} \label{fig1} }
\end{figure}

\clearpage
\begin{figure}
\includegraphics[width=12cm,angle=0]{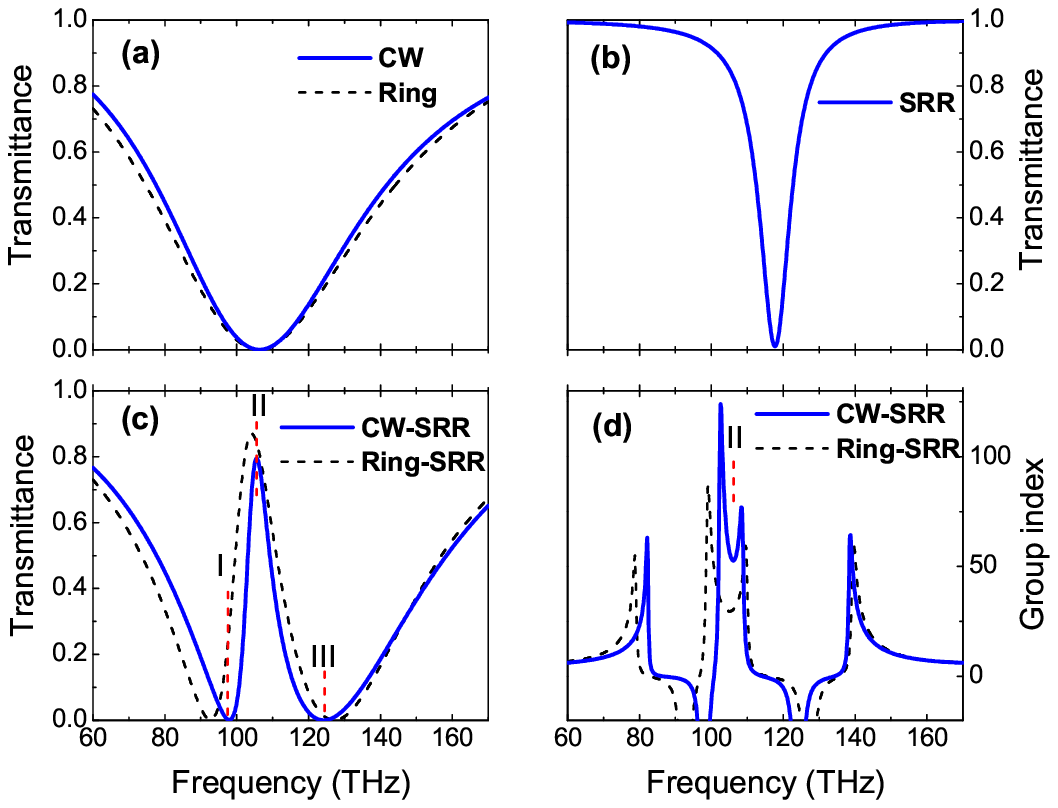}
\caption{Thuy \emph{et al.} \label{fig2} }
\end{figure}

\clearpage
\begin{figure}
\includegraphics[width=12cm,angle=0]{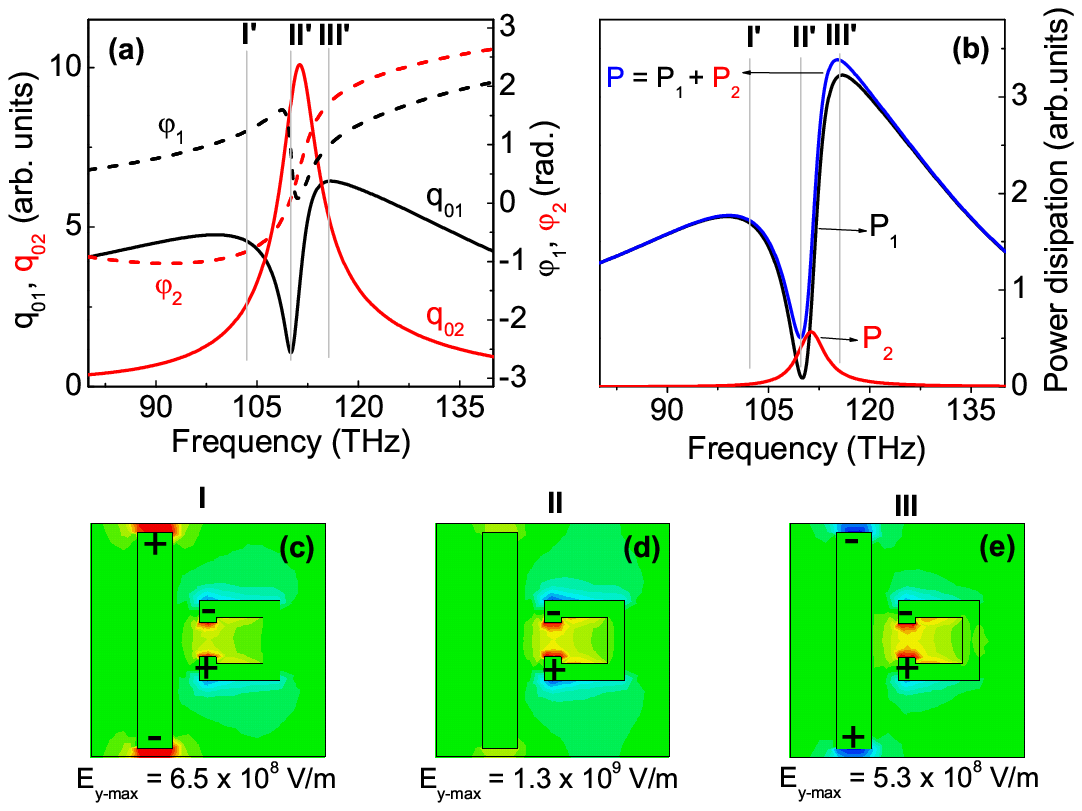}
\caption{Thuy \emph{et al.} \label{fig3} }
\end{figure}

\clearpage
\begin{figure}
\includegraphics[width=12cm,angle=0]{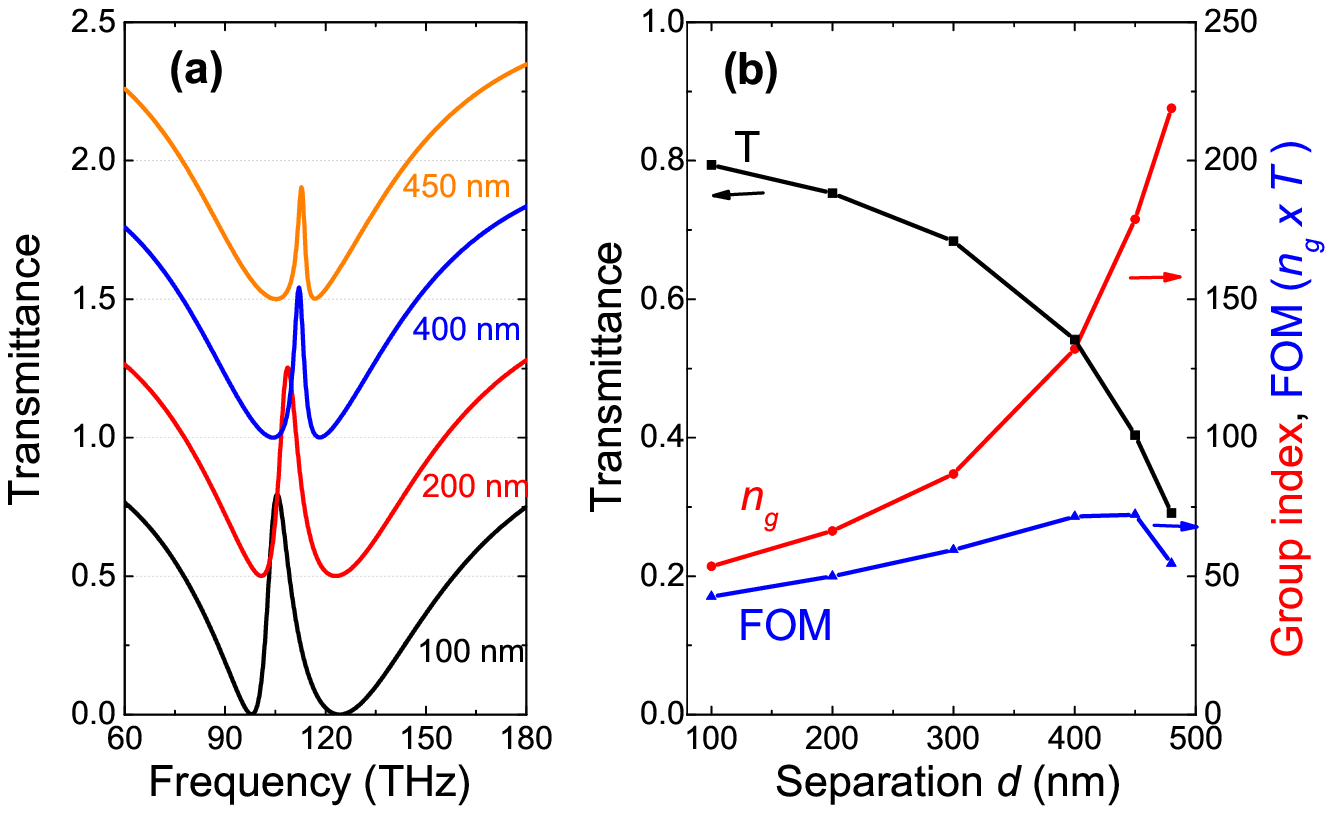}
\caption{Thuy \emph{et al.} \label{fig4} }
\end{figure}


\begin{thebibliography}{00}
\bibitem{22} Lumped values were roughly estimated as follows; from the geometry of SRR and CW we estimated $C_{2}\approx 0.04$ aF (see Ref. 23), and $C_{1}\approx4C_{2}$. The values of $L_{1,2}$ and $R_{1,2}$ are then calculated to match the resonance frequencies $f_{1,2}=\sqrt{1-R_{1,2}^{2}C_{1,2}/(4L_{1,2})}/\sqrt{L_{1,2}C_{1,2}}\approx10^{2}$ THz and the $Q$-factors ($Q_{1}=R_{1}^{-1}/\sqrt{L_{1}C_{1}}=1.6$ and $Q_{2}=R_{2}^{-1}/\sqrt{L_{2}C_{2}}=11.5$ for CW and SRR, repectively). We found that $L_{1}\approx 2.5$ pH, $R_{1}\approx 140$ $\Omega$, $L_{2}\approx 10$ pH and $R_{2}\approx 10$ $\Omega$, and took $C_{c}=0.12$ aF.
\end{thebibliography}
\end{document}